\documentclass[epj]{webofc}
\usepackage[utf8]{inputenc}
\usepackage[varg]{txfonts}   
\usepackage{booktabs}
\usepackage{xcolor}
\definecolor{darkred}{rgb}{0.4,0.0,0.0}
\definecolor{darkgreen}{rgb}{0.0,0.4,0.0}
\definecolor{darkblue}{rgb}{0.0,0.0,0.4}
\usepackage[bookmarks,linktocpage,colorlinks,
    linkcolor = darkred,
    urlcolor  = darkblue,
    citecolor = darkgreen]{hyperref}
\usepackage{subfigure,nicefrac,slashed, mciteplus}
\wocname{EPJ Web of Conferences}
\woctitle{Lattice2017}
\begin{document}
%
\selectlanguage{english}
\title{%
QCD at finite isospin chemical potential
}
\author{%
\firstname{Bastian B.} \lastname{Brandt}\inst{1}\fnsep\thanks{Speaker, \email{brandt@th.physik.uni-frankfurt.de}} \and
\firstname{Gergely} \lastname{Endr\H{o}di}\inst{1} \and
\firstname{Sebastian}  \lastname{Schmalzbauer}\inst{1}\fnsep\thanks{Speaker, \email{schmalzbauer@th.physik.uni-frankfurt.de}}
}
\institute{%
Institute for Theoretical Physics, Goethe Universit\"at Frankfurt, D-60438 Frankfurt am Main, Germany
}
\abstract{%
  We investigate the properties of QCD at finite isospin chemical potential
  at zero and non-zero temperatures. This theory is not affected by the sign problem
  and can be simulated using Monte-Carlo techniques. With increasing isospin
  chemical potential and temperatures below the deconfinement transition the
  system changes into a phase where charged pions condense, accompanied by
  an accumulation of low modes of the Dirac operator. The simulations
  are enabled by the introduction of a pionic source into the action, acting
  as an infrared regulator for the theory, and physical results are obtained
  by removing the regulator via an extrapolation. We present an update of our
  study concerning the associated phase diagram using 2+1 flavours of staggered
  fermions with physical quark masses and the comparison to Taylor expansion.
  We also present first results for our determination of the equation of state
  at finite isospin chemical potential and give an example for a cosmological
  application. The results can also be used to gain information about QCD
  at small baryon chemical potentials using reweighting with respect to
  the pionic source parameter and the chemical potential and we present first
  steps in this direction.
}
\maketitle
\section{Introduction}
\label{sec:intro}

The study of first principles QCD under extreme conditions is of vital
importance to illuminate the properties of QCD dominated aspects of matter
in the universe and for phenomenology to test and develop models which can be
used to understand different aspects of matter surrounding us.
Despite the advances and successes of simulating QCD on a lattice in the past
decades most regions of the QCD parameter space are still mainly unexplored.
Particularly challenging is the study of QCD at finite density, since it is
affected by the well-known sign problem, hindering numerical simulations.
In the sector of the light up ($u$) and down $(d)$ quarks the finite density
parameter space can be characterised by baryon, $\mu_B=(\mu_u+\mu_d)/2$,
and isospin, $\mu_I=(\mu_u-\mu_d)/2$, chemical potentials, as conjugate
parameters to the associated densities, $n_B$ and $n_I$, in the grand
canonical ensemble. The sign problem appears as soon the baryon chemical
potential is non-vanishing, while QCD at finite isospin chemical but with
$\mu_B=0$ is sign-problem-free and thus permits simulations of lattice QCD.

\begin{figure}[t]
  \centering
  \begin{minipage}{.48\textwidth}
  \centering
  \includegraphics[width=\textwidth,clip]{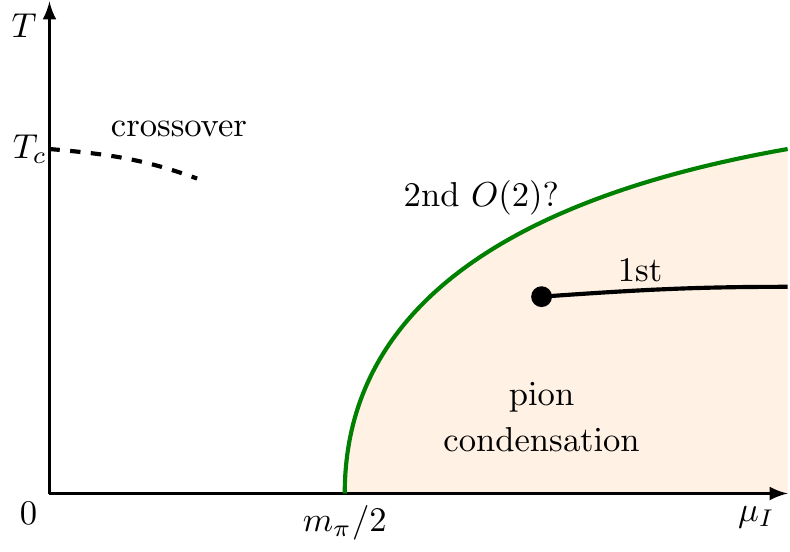}
  \end{minipage}
  \caption{Left: Schematic picture of the QCD phase diagram at finite isospin
  chemical potential, based on the findings in chiral perturbation
  theory~\cite{Son:2000xc}. The different phases are explained in the text.
  The black line indicates the conjuctured pure gauge deconfinement transition
  extending into the space of finite $\mu_I$ from the $\mu_I\to\infty$ limit.}
  \label{fig:phd-schem}
\end{figure}

While typical systems in nature involving strongly interacting matter, such as
the early universe, compact stars or heavy-ion collisions, share both,
non-vanishing baryon and isospin densities, the
study of QCD at pure isospin chemical potential is important and interesting
in its own right. The expected phase diagram based on the findings in chiral
perturbation theory~\cite{Son:2000xc}, depicted schematically in
fig.~\ref{fig:phd-schem}, is expected to develop phases similar to the ones
in the phase diagram at finite baryon chemical potential. In
particular, at zero temperature and small $\mu_I$ the system shows the
so-called Silver Blaze phenomenon~\cite{Cohen:2003kd}, where the groundstate
of the system is not affected by $\mu_I$. When the system crosses the
threshold chemical potential, $\mu_I=m_\pi/2$, charged pions can be created,
leading to pion condensation~\cite{Migdal:1978az,Ruck:1976zt}.
The associated phase transition is expected to be of second order in the
$O(2)$ universality class. While unimportant for most of the situations with
isospin asymmetric matter, pion condensation can potentially play an important
role in the description of neutron stars and for nuclear physics. Pion
condensation also goes hand-in-hand with a proliferation of low-modes of the
Dirac operator, leading to numerical problems in the simulations. A similar
accumulation of low modes is also expected at finite baryon chemical
potentials above threshold. In the pion condensation phase simulations
are only possible with the use of an infrared regulator in the form of
a pionic source term with parameter $\lambda$ which is introduced into the
action, see~\cite{Kogut:2002zg,Endrodi:2014lja}. Physical results are then
obtained by extrapolating the results to $\lambda\to0$, providing the main
challenge in the analysis step.

At small temperatures the above features of the phase diagram are expected
to remain mainly unaffected up to a possible shift of the pion condensation
phase boundary, see fig.~\ref{fig:phd-schem}. Around the chiral symmetry restoration/deconfinement transition temperature $T_c$,
the pion condensate is expected to `melt' or `evaporate'. Consequently, the
pion condensation phase boundary can potentially be shifted to very large
values of $\mu_I$ when $T>T_c$. It is then interesting to
investigate the interplay between the phase boundary to the pion condensation
phase and the chiral symmetry restoration crossover. For extremely large
values of $\mu_I$ a decoupling of the quark and gluon degrees of freedom and
a first-order deconfinement phase transition associated with the gluon sector
of the theory is expected to take place~\cite{Son:2000xc}. Thus it has been
concluded that there might be a second phase transition for large values of $\mu_I$,
depicted by the solid black line in fig.~\ref{fig:phd-schem}, which would then
end on a second order critical point~\cite{Son:2000xc}.

In the past decade a number of groups have investigated the properties of
QCD at finite isospin chemical potential on the lattice~\cite{Kogut:2002zg,Kogut:2002tm,Kogut:2004zg,deForcrand:2007uz,Detmold:2012wc,Endrodi:2014lja}
and in a variety of other approaches (see for example refs.~\cite{Splittorff:2000mm}).
However, all of the studies so far have
been done on coarse lattices using unphysically large pion masses
and/or an unphysical flavour content. In~\cite{Brandt:2016zdy} we have
presented first results of our study in the setup with 2+1 flavours
of stout-improved staggered fermions at physical quark masses.
In particular, we presented a novel method for the $\lambda\to0$
extrapolation using the singular values of the massive Dirac operator and
presented first results for our study of the phase diagram and the
comparison to Taylor expansion from $\mu=0$ on $N_t=6$ lattices. In this
proceedings article we update these results
by showing new results for $N_t=6$ and first results from $N_t=8$
and $10$ lattices. We also present first results for the measurement of
the equation of state at finite $\mu_I$.
The associated result for the
pressure at $T=0$ can be used to construct a hypothetical boson star
made of pions.
First results and prospects for such a construction will be presented in
section~\ref{sec:star}.
Finally, we show how our results can be used to obtain information about
the phase diagram of QCD at finite baryon chemical potential using
reweighting. A particularly interesting, and up to date mostly unanswered,
question concerns the properties of QCD in the enlarged $(\mu_B,\mu_I)$
parameter space. Our $\mu_B=0$ simulations are idealy suited to study this
parameter space for small values of $\mu_B$. First results in this
direction will be reported in section~\ref{sec-5}.

\section{Simulation setup and \texorpdfstring{\boldmath $\lambda$}{lambda}-extrapolations}
\label{sec-1}

In this section we sketch the setup of the lattice simulation. In particular, we show
how the pionic source term is included into the lattice
action, define relevant observables and discuss the methods used for the $\lambda\to0$
extrapolations. 

\subsection{Lattice action}
\label{sec:action}

We consider lattice QCD with three quark flavors $u,d,s$ at temperature $T=(aN_t)^{-1}$
in a discretized volume $V=(aN_s)^3$ with lattice spacing $a$. The quark masses $m_u=m_d=m_{ud}$ and $m_s$ are chosen to be physical and tuned along the line of constant physics~\cite{Aoki:2005vt,Borsanyi:2010cj}.
The partition function is given by a
path integral over all possible gauge fields,
\begin{equation}
	\label{eq:Z}
	Z_{\mu_I,\lambda} = \int \!\mathcal{D}[U]\; \left(\det
	\mathcal{M}_{ud}\right)^{\nicefrac{1}{4}} \left(\det \mathcal{M}_{s}\right)^{\nicefrac{1}{4}} e^{- \beta S_G},
\end{equation}
where $\beta$ is the inverse coupling, $S_G$ the Symanzik improved gauge action and
\begin{equation}
	\label{eq:quark-matrices}
	\mathcal{M}_{ud} = \begin{pmatrix}
		\slashed{D}(\mu_I) + m_{ud} & \lambda \eta_5 \\
		-\lambda \eta_5 & \slashed{D}(-\mu_I) + m_{ud}
	\end{pmatrix}
	\qquad
	\mathcal{M}_{s} = \slashed{D}(0) + m_s
\end{equation}
are the Dirac operators in the light and the strange sectors, respectively.
The quartic roots in eq.~(\ref{eq:Z}) originate from the rooting procedure to
remove the unwanted tastes from the simulations which appear due to the use of the staggered
Dirac operator $\slashed{D}(\mu)$. The off-diagonal elements in $\mathcal{M}_{ud}$
originate from the introduction of a pionic source term and break the residual $U_{\tau_3}(1)$
symmetry of the action at $\lambda=0$ explicitly.
$\eta_5 = (-1)^{n_x + n_y + n_z + n_t}$ is the staggered
fermion equivalent to $\gamma_5$. For $\lambda \in \mathbb{R}$ and $m_s > 0$, Monte-Carlo
techniques like RHMC can be applied directly because of
\begin{equation}
	\label{eq:positivity}
	\det \mathcal{M}_{ud} = \det \left(|\slashed{D}(\mu_I) + m_{ud}|^2 + \lambda^2 \right)
	> 0, \qquad \det \mathcal{M}_s = \det \left (|\slashed{D}(0) + m_s|^2\right) > 0,
\end{equation}
following from
\begin{equation}
	\label{eq:hermiticity}
	\eta_5 \tau_1 \mathcal{M}_{ud} \tau_1 \eta_5 = \mathcal{M}_{ud}^{\dagger},
	\qquad \eta_5 \mathcal{M} \eta_5 = \mathcal{M}_{s}^{\dagger}.
\end{equation}

\subsection{Observables and \texorpdfstring{\boldmath $\lambda$}{lambda}-extrapolations}
\label{sec:obs+lambda}

To extract information about the phase diagram, we study the pion condensate, the
light quark condensate and the isospin density,
\begin{equation}
	\label{eq:obs-def}
	\langle \pi \rangle = \frac{T}{V} \frac{\partial \ln Z}{\partial \lambda},\qquad
	\langle \bar{\psi} \psi \rangle = \frac{T}{V}\frac{\partial \ln Z}{\partial m_{ud}},\qquad
	\langle n_I \rangle = \frac{T}{V}\frac{\partial \ln Z}{\partial \mu_I}.
\end{equation}
Plugging (\ref{eq:Z}) into these definitions and making use of (\ref{eq:positivity}),
they are explicitly given by
\begin{equation}
	\label{eq:obs}
	\begin{aligned}
	\langle \pi \rangle &= \frac{T}{2V} \left \langle \mathrm{tr} \frac{\lambda}{|\slashed{D}(\mu_I)
	+ m_{ud}|^2 + \lambda^2} \right \rangle, \\
	\langle \bar{\psi} \psi \rangle &= \frac{T}{2V} \left \langle \Re \mathrm{tr} \frac{\slashed{D}(\mu_I)
	+ m_{ud}}{|\slashed{D}(\mu_I) + m_{ud}|^2 + \lambda^2} \right \rangle, \\
	\langle n_I \rangle &= \frac{T}{2V} \left \langle \Re \mathrm{tr} \frac{\left[\slashed{D}(\mu_I)
	+ m_{ud} \right]^{\dagger} \cdot \partial \slashed{D}(\mu_I)/\partial \mu_I}{|\slashed{D}(\mu_I) +
	m_{ud}|^2 + \lambda^2} \right \rangle \,.
	\end{aligned}
\end{equation}
The traces can be evaluated using stochastic estimators or in the basis of the singular
values of the massive Dirac operator~\cite{Brandt:2016zdy}.

$\langle \bar{\psi} \psi \rangle$ and $\langle \pi \rangle$ are subject to renormalisation
and, following~\cite{Brandt:2016zdy}, we define the renormalised condensates as
\begin{equation}
	\label{eq:renormalization}
	\Sigma_{\bar{\psi}\psi} = \frac{m_{ud}}{m_{\pi}^2 f_{\pi}^2}\left[\langle
	\bar{\psi}\psi\rangle_{T,\mu_I} - \langle \bar{\psi} \psi \rangle_{0,0} \right] +
	1, \qquad \Sigma_{\pi} = \frac{m_{ud}}{m_{\pi}^2 f_{\pi}^2} \langle \pi \rangle,
\end{equation}
where we have introduced the pion mass $m_{\pi} = 135~\mathrm{MeV}$ and the chiral limit of the pion decay
constant $f_{\pi}=86~\mathrm{MeV}$ for the purpose of normalisation.

Measurements of the observables are done for the ensemble including the artificial
pion source term proportional to $\lambda$. To obtain physical results it is thus
necessary to extrapolate the results to $\lambda=0$. This $\lambda\to0$ extrapolation
is the crucial and most difficult step concerning the analysis of the data. As shown
in~\cite{Brandt:2016zdy} a naive extrapolation is cumbersome and can lead to large
systematical uncertainties. An improvement program for the
$\lambda$-extrapolations has been outlined in~\cite{Brandt:2016zdy}, based on the
singular value representation of the operator traces mentioned above. The first step
consists of a ``valence quark improvement'', where the observable is replaced by
(an approximation of) its $\lambda=0$ counterpart. The simplest way to do this would
be to set $\lambda=0$ on the right hand sides of (\ref{eq:obs}). This, however, is
not possible in practice, due to the accumulation of small eigenvalues at finite
$\mu_I$, leading to problems with the inversions in the stochastic approximation
of the traces. Furthermore, for the pion condensate the source term is explicitly
needed to obtain a non-vanishing value. The singular value representation of the
traces in eq.~(\ref{eq:obs}) provides an alternative for the computation of the
traces and we have shown in~\cite{Brandt:2016zdy} how it can be used to reformulate
the pion condensate in terms of the density $\rho(\xi)$ of singular values of the
massive Dirac operator $\xi$ (see~\cite{Kanazawa:2011tt} for the derivation in
the massless case),
\begin{equation}
	\label{eq:pi_rho}
	\langle \pi \rangle = \pi \cdot \langle \rho(0) \rangle \,.
\end{equation}
A similar improvement can also be done for the other observables and the associated
publication containing the details is in preparation. This valence improvement
already removes most of the $\lambda$-dependence of the expectation value, as we
will show in Sec.~\ref{sec:lambda-rew}. The remaining $\lambda$-dependence can be
further reduced, constituting the second step in the improvement procedure, by
reweighting the resulting expectation value with the leading order
expansion of the full reweighting factor in $\lambda$. The details are
provided in~\cite{Brandt:2016zdy}. The remaining $\lambda$-dependence is mostly
flat and can be extrapolated to $\lambda=0$ in a well controlled manner.

\section{Thermodynamics at finite isospin chemical potential}
\label{sec-2}

In this section we will present our results regarding the phase
diagram, the equation of state and compare our results to the
ones from Taylor expansion around $\mu_I=0$. In the following
all of the results have already been extrapolated to $\lambda=0$
using the machinery described above. We will also briefly discuss
a possible cosmological application for the equation of state
determined at $T=0$.

\subsection{Results for the phase diagram}

We start by updating the results concerning
the phase diagram at finite $\mu_I$ shown in ref.~\cite{Brandt:2016zdy}.
In particular, we present new results for the crossover temperatures
for $\mu_I<m_\pi/2$ on the $N_t=6,8$ and 10 lattices and the
resulting phase diagram. We also discuss the location of the chiral
symmetry restoration transition within the pion condensation region.

The pseudocritical temperature of the crossover can be defined by the
inflection point of the renormalised condensate from eq.~(\ref{eq:obs}), for
instance. In terms of this definition the crossover temperature in the
continuum at $\mu_I=0$ is $T_c(0)=155(3)(3)$~MeV~\cite{Borsanyi:2010bp}.
In the present study we start with a slightly different definition and
determine $T_c(\mu_I)$ by requiring that $\Sigma_{\bar\psi\psi}$ takes
the value $\left.\Sigma_{\bar\psi\psi}\right|_{T_c}\approx-0.550$ at
$T_c(\mu_I)$, see~\cite{Brandt:2016zdy}. This value agrees with the
result for $\left.\Sigma_{\bar\psi\psi}\right|_{T_c}$ obtained in the
continuum limit at $\mu_I=0$~\cite{Borsanyi:2010bp}. Note, that this
definition is only valid as long as we are considering the crossover in
the Silver Blaze region, $\mu_I<m_\pi/2$, where the condensate at $T=0$
is independent of $\mu_I$, but does not hold for $\mu_I>m_\pi/2$.

\begin{figure}[t]
  \centering
  \includegraphics[width=\textwidth,clip]{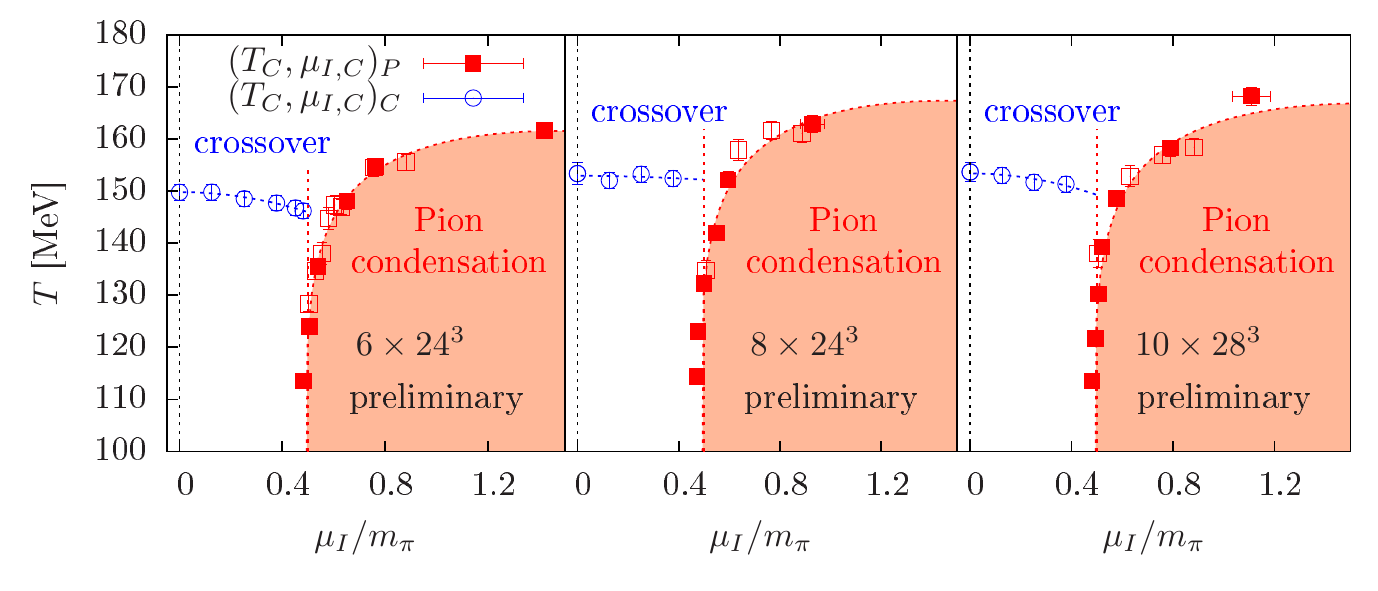}
  \caption{Phase diagram for the $6\times24^3$ (left), $8\times24^3$
  (middle) and $10\times28^3$ (right) lattices. The red squares are the
  results for the phase boundary to the pion condensation phase, 
  $(T_c,\mu_{I,c})_P$, and the blue points the ones  for the crossover
  line, $(T_c,\mu_{I,c})_C$. The open red squares have been obtained
  from scans in the temperature and the filled ones from scans in $\mu_I$.}
  \label{fig:phdiag-full}
\end{figure}

In figure~\ref{fig:phdiag-full} we show the resulting phase diagram for
$N_t=6,8$ and 10. The results for the pion condensation phase are those
from ref.~\cite{Brandt:2016zdy} and have been extracted from the
points where the system develops a non-zero pion condensate.
As observed already in ref.~\cite{Brandt:2016zdy} for $N_t=6$, the
crossover temperature at $\mu_I=0$ appears to lie somewhat below the
temperature associated with the melting of the pion condensate at high
chemical potentials. Both
temperatures tend to increase slightly in the approach to the continuum,
as expected for the crossover temperature at $\mu_I=0$, which should
approach $T_c(0)=155(3)(3)$~MeV, but the qualitative picture remains
unchanged. The chiral symmetry restoration transition, however, is a
broad crossover, whereas pion condensation sets in via a true phase
transition.The latter is supported by a finite size scaling study in
ref.~\cite{Brandt:2016zdy}. The crossover line
shows a downwards trend for growing $\mu_I$,
even though the associated curvature shows large
fluctuations for $a\to0$.

\begin{figure}[t]
  \centering
  \begin{minipage}{.48\textwidth}
  \centering
  \includegraphics[width=\textwidth,clip]{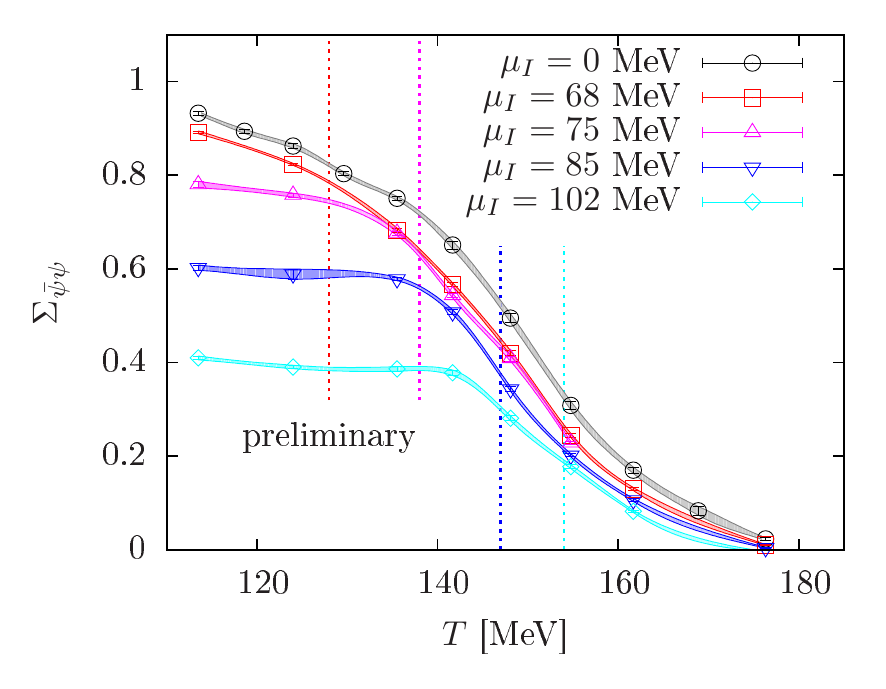}
  \end{minipage}
  \caption{Results for $\Sigma_{\bar\psi\psi}$ vs. the temperature for
  different values of $\mu_I>m_\pi/2$ compared to the data at $\mu_I=0$. 
  The colored bands are the results from a cubic spline interpolation and the
  dashed vertical lines indicate the pion condensation phase
  boundary for the value of $\mu_I$ associated to the same color.}
  \label{fig:chcond-pcp}
\end{figure}

To answer the question whether the chiral symmetry restoration transition
and the phase boundary of the pion condensation phase start to coincide
for $\mu_I>m_\pi/2$ we need to change the definition of the crossover
temperature and determine $T_c$ via the inflection point of the condensate.
While this is still work in progress (and will eventually also replace the
definition for $T_c$ for $\mu_I<m_\pi/2$), a look at the behaviour of the
condensate for $\mu_I>m_\pi/2$ can be suggestive for the results that one
might expect. We show the condensate for different values of
$\mu_I>m_\pi/2$ on the $N_t=6$ lattice in fig.~\ref{fig:chcond-pcp} in
comparison to the condensate at $\mu_I=0$.
The different starting points of the condensate at $T=113$~MeV are due
to the fact that the $T=0$ condensate changes its value in the pion
condensation phase, i.e. for $\mu_I>m_\pi/2$. Consequently the
subtraction of the $T=0$ and $\mu=0$ condensate -- needed for the renormalization (7) -- leads
to curves that do not start at unity. The dashed vertical lines
indicate the temperature of the pion condensation phase boundary for these
values of $\mu_I$. There is a clearly visible trend that the inflection
point of the condensate and the pion condensation phase boundary approach
each other with increasing $\mu_I$, meaning that one would expect chiral
symmetry restoration to set in at the boundary of the pion condensation
phase.

\subsection{Testing Taylor expansion}

As outlined in the introduction, one of the main challenges for simulations
in lattice QCD is the sign problem for non-zero baryon chemical potential
$\mu_B$. For small values of $\mu_B$, the sign problem can be overcome
either by reweighting (see section~\ref{sec:mu-rew}) or via the Taylor expansion
method. In the latter, expectation values of observables are expanded around
$\mu_B=0$. The resulting expressions contain the derivatives of the observable
with respect to $\mu_B$ evaluated at $\mu_B=0$, which can be computed numerically.
The main problem of the method is the {\it a priori} unknown range of applicability 
for a fixed order of the expansion. A similar Taylor expansion can also be
performed for non-zero isospin chemical potentials, so that our results
can be used to explicitly check the range of applicability of the method.

As before (see ref.~\cite{Brandt:2016zdy}) we will focus on the isospin
density $n_I$ for which the Taylor expansion with respect to $\mu_I$ is
given by
\begin{equation}
\frac{\left\langle n_I \right\rangle}{T^3} = c_2 \Big(\frac{\mu_I}{T}\Big) +
\frac{c_4}{6} \Big(\frac{\mu_I}{T}\Big)^3 \,,
\end{equation}
with the Taylor coefficients $c_2$ and $c_4$ (the expressions for $c_2$ and
$c_4$ are provided in ref.~\cite{Brandt:2016zdy}). To determine $c_2$ and
$c_4$ we use a cubic spline interpolation of the Taylor expansion coefficients
from ref.~\cite{Borsanyi:2011sw}, where the same action has been used.

\begin{figure}[t]
  \centering
  \begin{minipage}{.48\textwidth}
  \centering
  \includegraphics[width=\textwidth,clip]{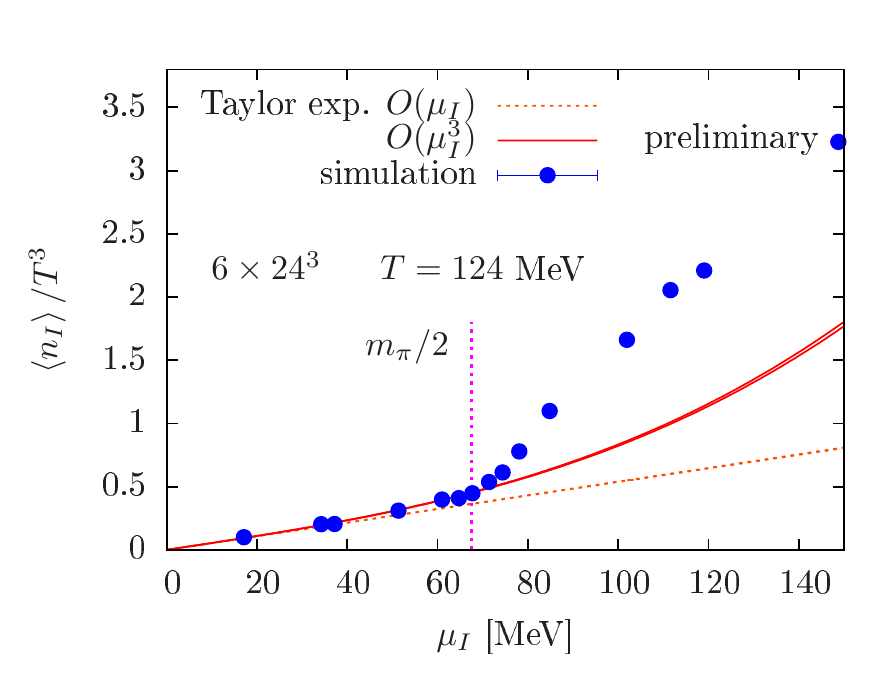}
  \end{minipage}
  \begin{minipage}{.48\textwidth}
  \centering
  \includegraphics[width=\textwidth,clip]{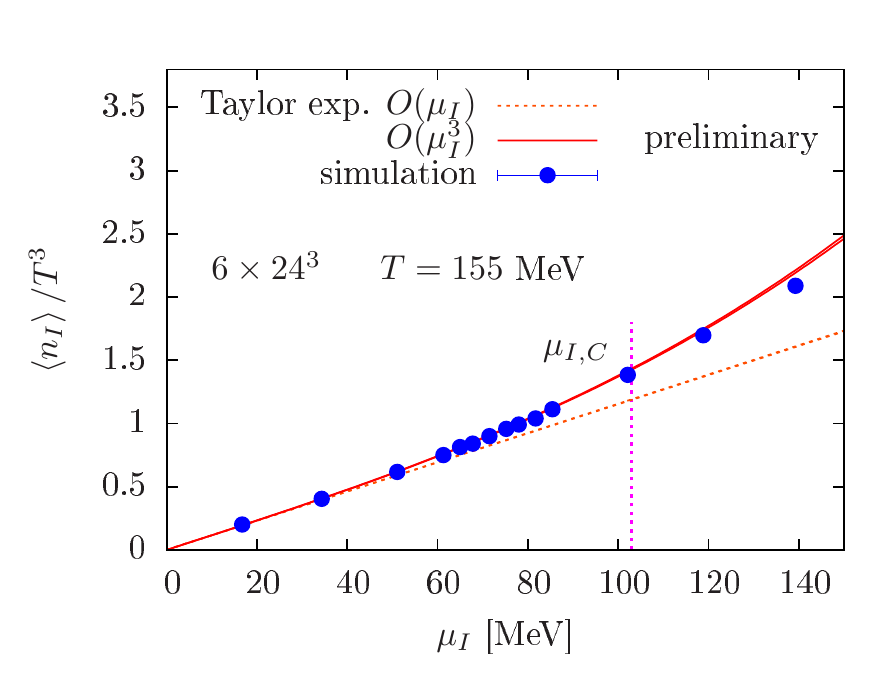}
  \end{minipage}
  \caption{Comparison of the results for $\left\langle n_I \right\rangle$
  from $6\times24^3$ lattices for temperatures 124 (left) and 155 MeV (right)
  and the results from Taylor expansion around $\mu_I=0$ to $O(\mu_I)$ and
  $O(\mu_I^3)$. The dashed vertical lines indicate the phase boundary to
  the pion condensation phase.
  \vspace*{-0.2cm}}
  \label{fig:nI-taylor}
\end{figure}

\begin{figure}[t]
  \centering
  \begin{minipage}{.48\textwidth}
  \centering
  \includegraphics[width=\textwidth,clip]{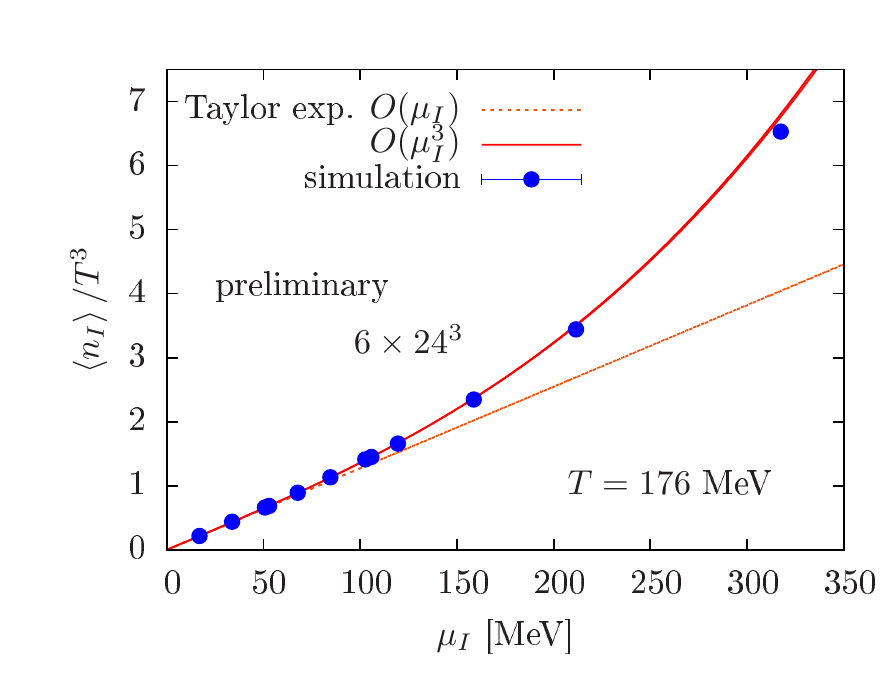}
  \end{minipage}
  \begin{minipage}{.48\textwidth}
  \centering
  \includegraphics[width=\textwidth,clip]{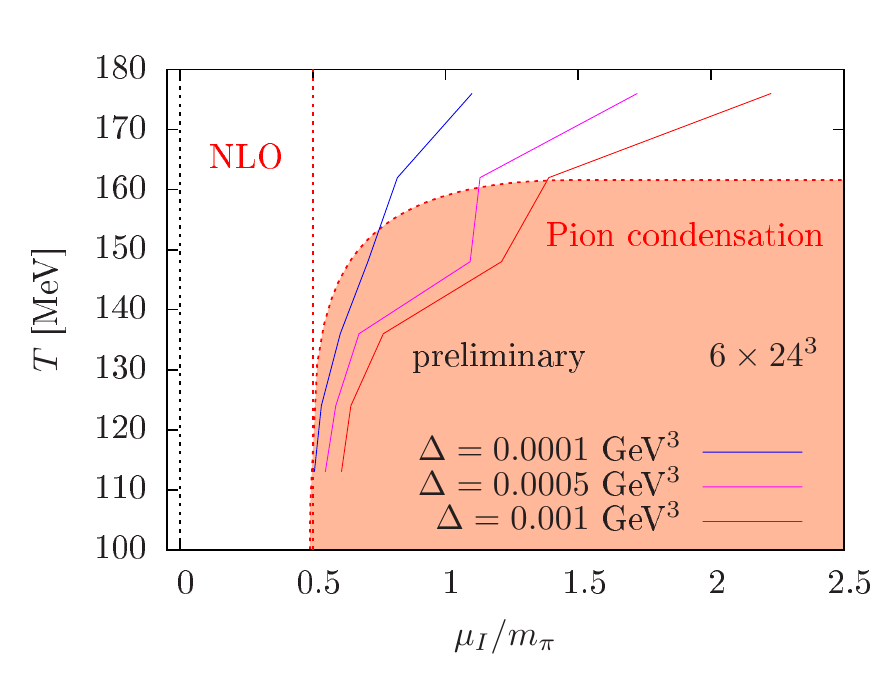}
  \end{minipage}
  \caption{Left: Same as in figure~\ref{fig:nI-taylor} for $T=176$~MeV.
  Note the difference in scale and that there is no vertical line since
  we are not entering the pion condensation phase for this temperature.
  Right: Contour plot of the difference $\Delta$ between the simulation
  results and Taylor expansion (see text). To generate the data for the
  plot, $\left\langle n_I \right\rangle$ has been interpolated using a
  cubic spline (see also section~\ref{sec:eos}).}
  \label{fig:nI-taylor2}
\end{figure}

In figure~\ref{fig:nI-taylor} we show the results for the comparison for
the $N_t=6$ lattice and two values of the temperature where we still reach
within the pion condensation phase. Note that by construction, the Taylor
expansion is expected to break down at the phase boundary. Indeed this
effect is clearly visible for $T=124$~MeV, while for $T=155$~MeV the
disagreement is less obvious. This may also be due to the fact that the
data remains in the vicinity of the phase boundary due to the strong
flattening visible in fig.~\ref{fig:phdiag-full}.
For both temperatures the data clearly follows the curves obtained from
Taylor expansion to $O(\mu_I^3)$, which for both temperatures becomes
distinct from the curve obtained from $O(\mu_I)$ between $\mu_I=50$ to
60~MeV with the present accuracy of the data (for $T=124$~MeV the difference
is only visible thanks to the improved $\lambda$-extrapolations,
which have improved the accuracy for $\left\langle n_I \right\rangle$
compared to the results presented in ref.~\cite{Brandt:2016zdy}).
In the left panel of figure~\ref{fig:nI-taylor2} we show a similar comparison
for $T=176$~MeV, where we do not enter the pion condensation phase. From
this plot we can see that the good agreement with Taylor expansion to
$O(\mu_I^3)$ extends all the way up to $\mu_I=200$~MeV, at least.
After that slight deviations seem to appear, signalling the importance
of terms of $O(\mu_I^5)$. To make this more quantifiable we show the lines
of constant difference $\Delta=|\left\langle n_I \right\rangle
-\left\langle n_I \right\rangle^{\rm Taylor}_{\rm NLO}|$, where the second
expectation value is the one obtained from Taylor expansion to $O(\mu_I^3)$,
in the right panel of figure~\ref{fig:nI-taylor2}. The plot shows that, as
expected, the data disagrees with the Taylor expansion as soon as one
enters the pion condensation phase. Above the phase boundary to the pion
condensation phase the good agreement of the data with the Taylor expansion
at a fixed order (here $O(\mu_I^3)$) extends to even larger values of $\mu_I$
for larger values of $T$, consistent with the notion that Taylor expansion
is actually an expansion in $\mu_I/T$. Apart from the results at $N_t=6$
shown here, results for $N_t=8$ are
also available and show no significant deviation from the findings above.
A more detailed study of the approach to the continuum is postponed to
a forthcoming publication.

\subsection{The equation of state}
\label{sec:eos}

One of the main ingredients for theoretical studies of phenomena and
objects in cosmology and nuclear physics is the QCD equation of state.
It is used for the hydrodynamic modelling of heavy-ion collisions and
for the construction of neutron stars, for instance, just to name two
of its many applications. For these physical situations the main 
contribution to the equation of state comes from finite baryon chemical
potential. Nonetheless, the equation of state is also affected by the
presence of a finite isospin chemical potential, so that the
associated effects should also be included for a complete description
of systems with isospin asymmetry.

Here we focus on the equation of state at pure isospin chemical
potential. The two main quantities we consider, and from which all of
the other quantities can be computed, are the pressure
\begin{equation}
 \frac{p}{T^4} = \frac{1}{VT^3} \log Z
\end{equation}
and the trace anomaly
\begin{equation}
 \frac{I}{T^4} = \frac{\epsilon - 3p}{T^4} = T
 \frac{\partial}{\partial T} \frac{p}{T^4} + \frac{\mu_I n_I}{T^4} \,.
\end{equation}
In this proceedings article we will show first results for the
pressure obtained from the $N_t=6$ lattices, while the computation
of the trace anomaly and other observables, as well as the continuum
limit are left for future publications.

\begin{figure}[t]
  \centering
  \begin{minipage}{.48\textwidth}
  \centering
  \includegraphics[width=\textwidth,clip]{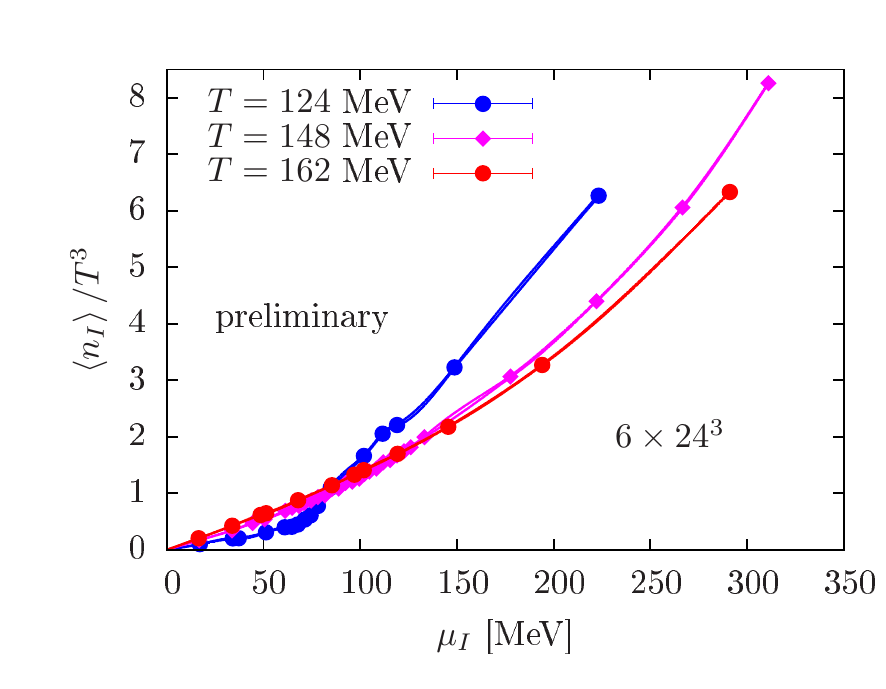}
  \end{minipage}
  \begin{minipage}{.48\textwidth}
  \centering
  \includegraphics[width=\textwidth,clip]{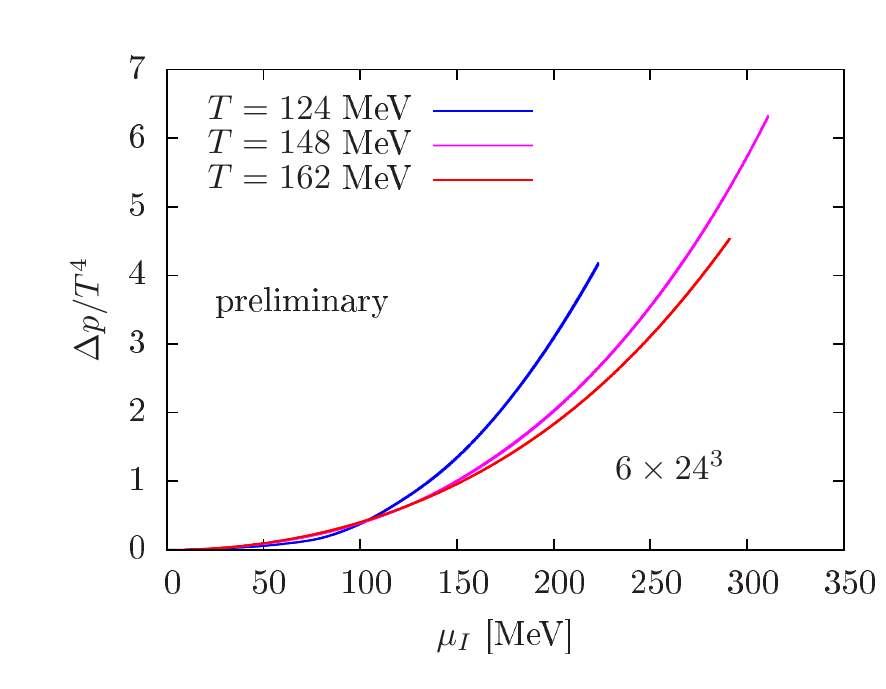}
  \end{minipage}
  \caption{Left: Spline interpolations for
  $\left\langle n_I \right\rangle$ used for the computation of the
  pressure. Right: Difference of the pressure at finite $\mu_I$ and
  the pressure at $\mu_I=0$, $\Delta p(T,\mu_I)$ from
  eq.~(\ref{eq:pint}) for different values of the temperature on the
  $6\times24^3$ lattices.}
  \label{fig:eos}
\end{figure}

The pressure can be rewritten as
\begin{equation}
\label{eq:pint}
 p(T,\mu_I) = p(T,0) + \int_0^{\mu_I} d\mu'_I \, n_I(T,\mu'_I)
 \equiv p(T,0) + \Delta p(T,\mu_I) \,,
\end{equation}
where we made use of the fact that $n_I=\partial p/(\partial \mu_I)$.
The main task for the computation of $p(T,\mu_I)$ is thus the
computation of $\Delta p(T,\mu_I)$, while $p(T,0)$ is known
from the interpolation provided in~\cite{Borsanyi:2010cj},
for instance. To compute $\Delta p(T,\mu_I)$ we evaluate the integral
using a cubic spline interpolation of the data for
$\left\langle n_I \right\rangle$, shown for some temperatures in the
left panel of fig.~\ref{fig:eos}. The results for
$\Delta p(T,\mu_I)$ are shown in the right panel of fig.~\ref{fig:eos}.
We can see that switching on $\mu_I$ leads to a general rise of the
pressure which is, not surprisingly, stronger for low temperatures,
where one enters the pion condensation phase.

\subsection{An application: Pion stars}
\label{sec:star}

The availability of a first-principles hadronic equation of state
is of tremendous importance in astrophysics, because it allows
to study compact stars under realistic conditions.
The phenomenon of pion condensation at $T=0$ and thus finite values
of the isospin density for $\mu_I>m_\pi/2$ could, at least in
principle, lead to the formation of cold and self-bound stars,
consisting of, say, positively charged
pions\footnote{For a similar proposal about pion stars and possible
production mechanisms see ref.~\cite{Carignano:2016lxe}.}.
These stars would correspond to a
class of hypothetical objects called boson
stars~\cite{Kaup:1968zz}\footnote{For reviews about studies
on boson stars see~\cite{Jetzer:1991jr,Liebling:2012fv}, for instance.}.
They are expected to be stable against pion decay, due to the massless
nature of the pions in the condensate.
In this section, we show how to construct such pion stars with the
equation of state obtained from our lattice 
simulations\footnote{For a similar study in $G_2$-QCD
see~\cite{Hajizadeh:2017jsw}.}. Since we
are considering a cold star we should use the equation of state
obtained at $T=0$. For this case measurements with the desirable
setup of physical quark masses and improved actions used above are
still in progress. Here we use the data for the
equation of state obtained from the setup used for the tests in
the next section instead.

The mass-radius relation of a static, spherically symmetric and
relativistic star can be obtained by solving the
Tolman-Oppenheimer-Volkoff (TOV) equation~\cite{norman1997compact}
\begin{equation}
\label{eq:TOV}
	\frac{\mathrm{d} p(r)}{\mathrm{d} r} = -G\frac{\left[p(r) +
	\epsilon(r)\right]\left[M(r) + 4\pi r^3 p(r)\right]}{r\left[r
	- 2GM(r)\right]} \,,
\end{equation}
with the gravitational constant $G = 6.70861(31)\cdot
10^{-39}~(\mathrm{GeV})^{-2}$, shell radius $r$, pressure $p(r)$,
energy density $\epsilon(r)$ and mass
\begin{equation}
\label{eq:M}
	M(r) = 4 \pi \int_{0}^{r} \!\mathrm{d} r' \; r'^2 \epsilon (r')\,.
\end{equation}
For a known relation $\epsilon(p)$, eqs. (\ref{eq:TOV}) and (\ref{eq:M}) can be 
solved simultaneously by numerical integration, starting from some central 
pressure $p(0)$ up to the edge of the star at radius $r=R$, indicated by a 
vanishing pressure $p(R)=0$. In our case, all necessary information is contained in
$\left\langle n_I \right\rangle$ since at
$T=0$
\begin{align}
	\label{eq:p(r)}
	p(r) = \int_{m_{\pi}/2}^{\mu_I(r)} \!\mathrm{d} \mu_I' \;
	\left\langle n_I \right\rangle(\mu_I') \quad \text{and} \quad
	\epsilon(r) = -p(r) + \mu_I(r) \langle n_I(\mu_I(r)) \rangle \,.
\end{align}
Eq.~(\ref{eq:TOV}) can then be rewritten as 
\begin{equation}
\label{eq:TOV_mu}
	\frac{\mathrm{d} \mu_I(r)}{\mathrm{d} r} = -G\mu_I(r)\frac{M(r) +
	4\pi r^3 p(r)}{r^2-2rGM(r)} \,,
\end{equation}
and the solutions may be labelled by the central chemical potential $\mu_I(0)$.
To obtain an interpolation for $\langle n_I(\mu_I) \rangle$ which
smoothly goes to zero at $\mu_I=m_\pi/2$, we use a cubic spline
interpolation of our lattice
data matched to chiral perturbation theory (see~\cite{Endrodi:2014lja})
around $\mu_I = m_{\pi}/2$.

\begin{figure}[t]
	\centering
        \begin{minipage}{.48\textwidth}
        \centering
	\includegraphics[width=\textwidth,clip]{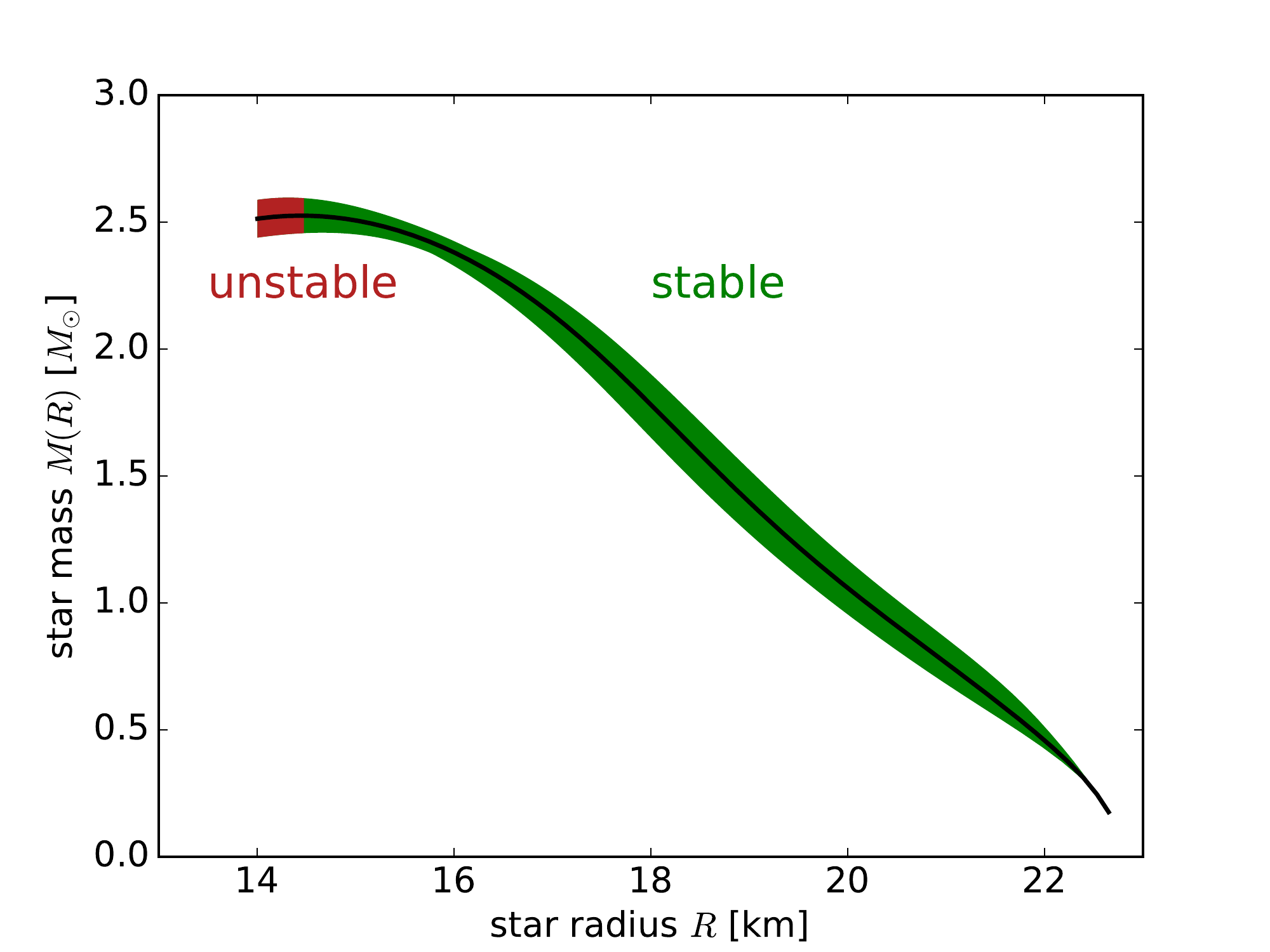}
	\end{minipage}
	\caption{Mass-radius relation of a pion star obtained with the TOV
	equation. The band includes both systematic and statistical errors. The stability was determined
	by checking the response to density fluctuations and radial
	oscillations~\cite{norman1997compact}. It is common to specify the
	results in astronomically meaningful units, i.e.
	$[R] =\mathrm{km}$ and $[M(R)] = M_{\odot} =
	1.116 \cdot 10^{57}~\mathrm{GeV}$.}
	\label{fig:M-R}
\end{figure}

The mass-radius relation obtained by solving
eq.~(\ref{eq:TOV}) for different initial values $\mu_I(0)$ is shown in
fig.~\ref{fig:M-R}. The masses of the resulting stars range up to around three
solar masses, similar to the expected masses of
neutron stars. At the same time the pion stars have a diameter which is
slightly larger than for neutron stars. Note, however, that these
results have been obtained with an equation of state describing positively
charged pions. This leaves us with a highly charged star, which is unlikely
to be stable when the effect of the electromagnetic interactions is included.
Neutrality can be reinstated by including further charged particles and
we are currently studying the resulting system.

\section{Reweighting to \texorpdfstring{\boldmath $\mu_B\neq0$}{non-zero baryon chemical potential}}
\label{sec-5}

\begin{figure}[t]
	\centering
	\includegraphics[width=0.4\textwidth,clip]{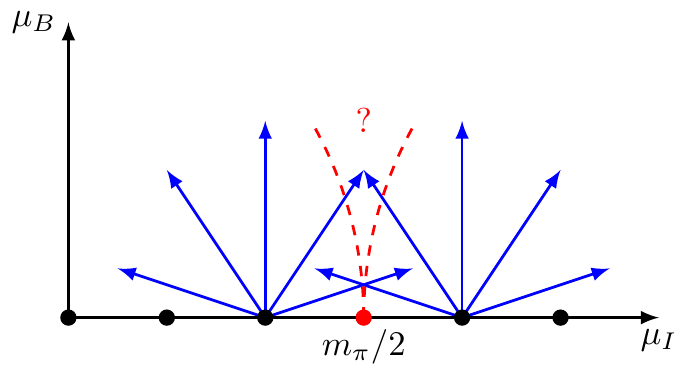}
	\caption{Sketch of how to obtain information about the phase boundary curvature
	(red dashed lines) in $\mu_B$ direction from pure $\mu_I$ simulation points
	(black dots) by reweighting (blue arrows).}
	\label{fig:muB-phase-boundary}
\end{figure}

Up to now, our investigations were restricted to vanishing baryon chemical
potential $\mu_B$. In most of the systems of interest, however, the finite
value of $\mu_B$ plays a crucial role. Since direct simulations at finite
$\mu_B$ are hindered by the sign problem, one can only use indirect methods
such as Taylor expansion or reweighting to obtain information about the
enlarged phase diagram in the $\mu_B-\mu_I$ plane. These methods are,
in general, restricted to small values of $\mu_B$ (for reweighting this
is due to the well-known overlap problem),
so that the phenomenologically interesting region beyond
the threshold of nucleon production cannot be reached. Apart from this
region at large $\mu_B$ the enlarged $\mu_B-\mu_I$ parameter space offers
other conceptually interesting regions, such as the region around the
$\mu_B=0$ axis for finite $\mu_I$, where it is interesting to investigate
the behaviour of the boundary to the pion condensation phase at finite
$\mu_B$, see fig.~\ref{fig:muB-phase-boundary}.

Since the simulations include a pion source, i.e. are performed at finite
values of $\lambda$, it is also necessary to reweight the data in $\lambda$
apart from the reweighting in the chemical potential. In this section we
introduce the methods used for both types of reweighting and present first
preliminary results. For testing purpose we switch to a cheaper setup which
has already been used in ref.~\cite{Endrodi:2014lja}. The results shown
here are obtained on $8^4$ lattices for $T\approx 0$, neglecting the
$s$-quark contribution, implementing unimproved staggered fermions and
using the simple Wilson plaquette gauge action. The lattice
spacing is $a=0.299(2)~\mathrm{fm}$ and we employ bare quark masses of
$am_{ud}=0.025$, resulting in a pion mass of about 260 MeV. Pion condensation sets in
at around $a\mu_I = am_{\pi}/2 \approx 0.2$.

\subsection{Reweighting in \texorpdfstring{\boldmath $\lambda$}{lambda}}
\label{sec:lambda-rew}

\begin{figure}[t]
        \begin{minipage}{.48\textwidth}
        \centering
	\includegraphics[width=\textwidth,clip]{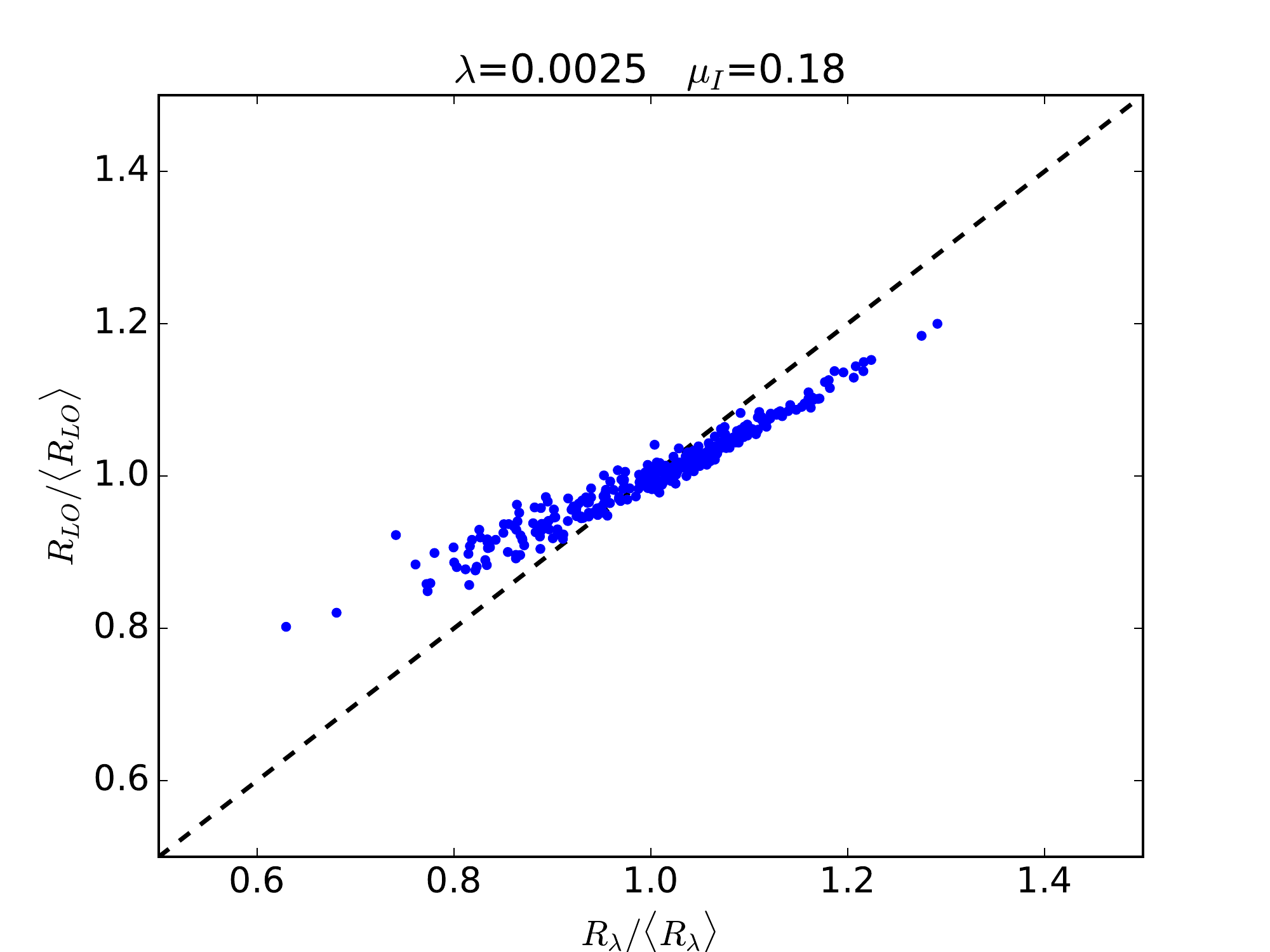}
	\end{minipage}
        \begin{minipage}{.48\textwidth}
        \centering
	\includegraphics[width=\textwidth,clip]{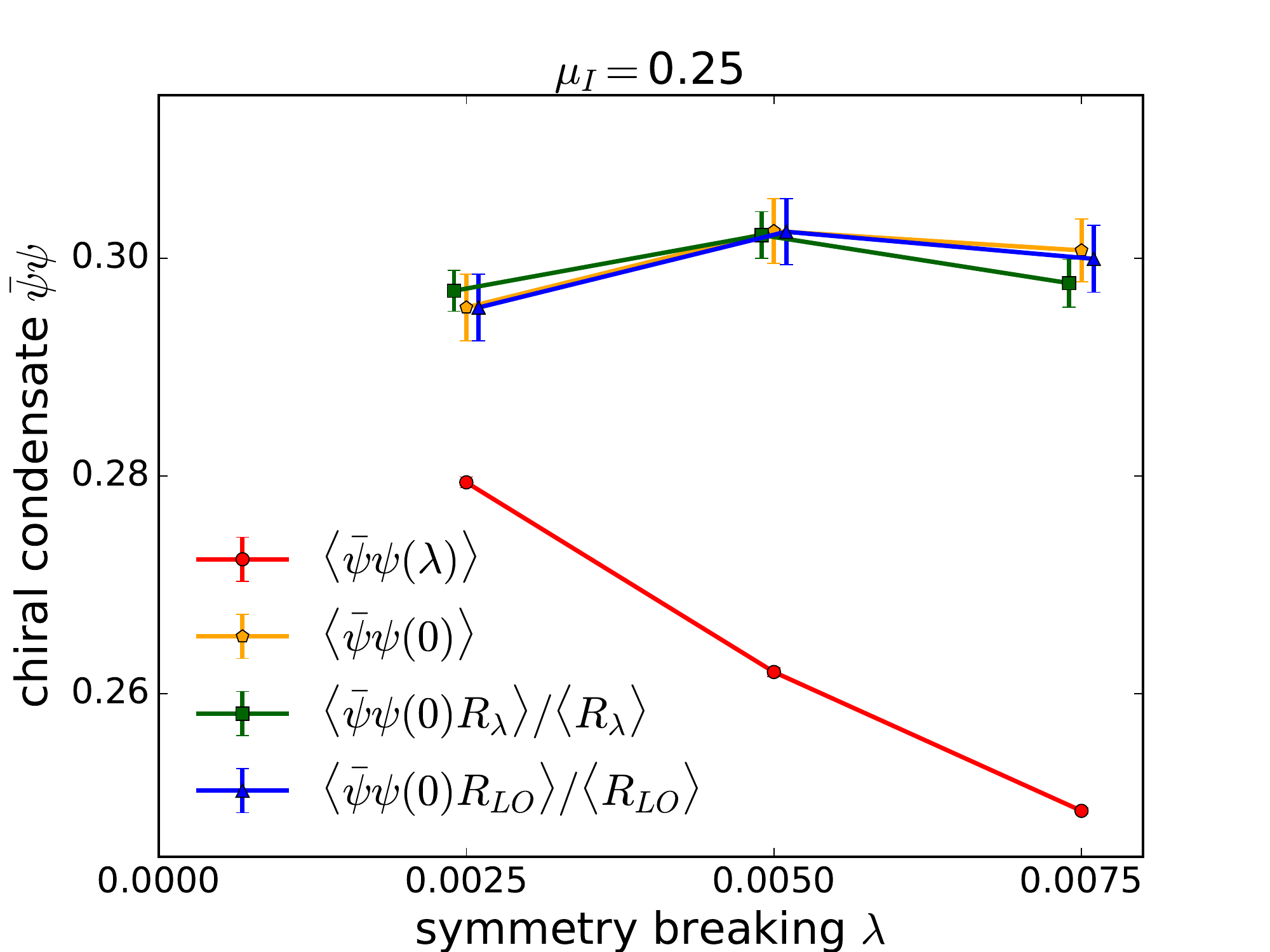}
	\end{minipage}
	\caption{Left: example of correlation between the full and leading-order reweighting factor, together with an indicated $R_{LO} = R_{\lambda}$ line. Right: comparison of full (green) and leading order (blue) reweighting in $\lambda$ with a naive (red) and improved (yellow) $\lambda$-extrapolation.\vspace*{-0.0cm}}
	\label{fig:l-extrapolations}
\end{figure}

The basic idea behind reweighting is the following. The expectation value of
an observable in a target ensemble, for us the pure isospin ensemble at
$\lambda=0$ with partition function
\begin{equation}
	\label{eq:Z_pure_isospin}
	Z_{\mu_I} = \int \!\mathcal{D}[U]\;  \det (M^{\dagger} M )^{\nicefrac{1}{4}}e^{-\beta S_G},
\end{equation}
is rewritten as a reweighted expectation value in an auxiliary ensemble.
Here the auxiliary ensemble includes the pionic source $\lambda$, so that
\begin{equation}
	\label{eq:rew-lambda}
	\begin{aligned}
	\langle O \rangle_{\mu_I} &= \frac{\int \!\mathcal{D}[U]\; \det(M^{\dagger}M)^{\nicefrac{1}{4}}e^{-\beta S_G} O}{\int \!\mathcal{D}[U]\; \mathrm{det}(M^{\dagger}M)^{\nicefrac{1}{4}} e^{-\beta S_G}} \\
	&= \frac{\int \!\mathcal{D}[U]\; \det(M^{\dagger}M + \lambda^2)^{\nicefrac{1}{4}} e^{-\beta S_G} R_{\lambda} O}{\int \!\mathcal{D}[U]\; \mathrm{det}(M^{\dagger}M + \lambda^2)^{\nicefrac{1}{4}} e^{-\beta S_G} R_{\lambda}} \times \frac{\int \!\mathcal{D}[U]\; \mathrm{det}(M^{\dagger}M + \lambda^2)^{\nicefrac{1}{4}} e^{-\beta S_G}}{\int \!\mathcal{D}[U]\; \mathrm{det}(M^{\dagger}M + \lambda^2)^{\nicefrac{1}{4}} e^{- \beta S_G}} \\
	&= \frac{\langle R_{\lambda} O\rangle_{\mu_I, \lambda}}{\langle R_{\lambda} \rangle_{\mu_I, \lambda}} \,,
	\end{aligned}
\end{equation}
with
\begin{equation}
	\label{eq:Rf}
	R_{\lambda} = \left[\frac{\mathrm{det}(M^{\dagger}M)}{\mathrm{det}(M^{\dagger}M + \lambda^2)}\right]^{\nicefrac{1}{4}} \,.
\end{equation}
Above we have used $M = \slashed{D}(\mu_I) + m_{ud}$ as a short notation.
The computational cost for the determination of the reweighting factors is
immense, as we have to compute all singular values of the massive Dirac
operator. The cost can potentially be reduced by using the leading order
expansion of the reweighting factor (cf. ref.~\cite{Brandt:2016zdy})
\begin{equation}
	\ln R_{\lambda} = \ln R_{LO} + \mathcal{O}\left(\lambda^4\right),
	\qquad \ln R_{LO}= -\frac{\lambda T}{2V}\pi \,.
\end{equation}
Here the pion condensate can be computed using stochastic estimators to reduce
the computational effort. Obviously, using only the leading order term as the
reweighting factor is an approximation and one needs to ensure that the associated
systematic effect is below the statistical uncertainty. One of the major
prerequisites to achieve this goal is a strong correlation between the two
types of reweighting factors. In the left panel of fig.~\ref{fig:l-extrapolations}
we show a scatter plot of the results for the normalised full and leading order
reweighting factors. The correlation between $R_{\lambda}$ and $R_{LO}$ is
clearly visible in the plot, but we observed that independent of $\lambda$,
there is a slight tilt in the correlation with respect to the $R_{\lambda}=R_{LO}$
line, which becomes more severe the bigger $\mu_I$. This behaviour can be explained
by the different response of the two reweighting factors to fluctuations of
eigenvalues. The details will be discussed in an upcoming paper.
In the right panel of fig.~\ref{fig:l-extrapolations} we show the effect of the
leading order reweighting compared to the full reweighting for the example of
the chiral condensate for different values of $\lambda$. Also included in the
plot are the results for the condensate evaluated at finite value of $\lambda$
and those without reweighting, but the operator evaluated at $\lambda=0$, as
explained in sec.~\ref{sec:obs+lambda}. The plot indicates that the systematic
effect associated with the use of the leading order reweighting compared to
the full reweighting factor is indeed below the uncertainties. In fact, once
we have improved the operator the remaining effect is small. In addition, we
see that for the shown values of $\lambda$ we do not observe any overlap problem
with the $\lambda=0$ ensemble, since the reweighting from different values
of $\lambda$ leads to results which agree within uncertainties. The green
squares correspond to the improved $\lambda$-extrapolations, discussed in
sec.~\ref{sec:obs+lambda}, and we can see that the associated $\lambda\to0$
extrapolation is flat and well under control.

\subsection{Reweighting in \texorpdfstring{\boldmath $\mu$}{the chemical potential}}
\label{sec:mu-rew}

To reach into the $\mu_B-\mu_I$ parameter space we still need to include the
reweighting step with respect to the chemical potentials of the light quarks
$\mu_u$ and $\mu_d$. The partition function of QCD with two mass degenerate
staggered quarks at arbitrary chemical potentials is given by
\begin{equation}
	\label{eq:Z-mu}
	Z_{\mu_u,\mu_d} = \int \!\mathcal{D}[U]\; \left[ \det M(\mu_u)\det M(\mu_d)\right]^{\nicefrac{1}{4}}e^{-\beta S_G}\,.
\end{equation}
Note that the factorization of the quark matrices is only possible due to the
absence of a pionic source. Once more we can rewrite expectation values in this target
ensemble as reweighted expectation values in the pure isospin ensemble
with $\lambda=0$ as
\begin{equation}
	\label{eq:rew-mu}
	\begin{aligned}
	\langle O \rangle_{\mu_u,\mu_d} &= \frac{\int \!\mathcal{D}[U]\;  \left[ \det M(\mu_u)\det M(\mu_d)\right]^{\nicefrac{1}{4}}e^{-\beta S_G} O}{\int \!\mathcal{D}[U]\;  \left[ \det M(\mu_u)\det M(\mu_d)\right]^{\nicefrac{1}{4}}e^{-\beta S_G}} \\
	&= \frac{\int \!\mathcal{D}[U]\;  \left[ \det M(\mu_I)\det M(-\mu_I)\right]^{\nicefrac{1}{4}}e^{-\beta S_G} R_{\mu} O}{\int \!\mathcal{D}[U]\;  \left[ \det M(\mu_I)\det M(-\mu_I)\right]^{\nicefrac{1}{4}}e^{-\beta S_G} R_{\mu}} \times \frac{\int \!\mathcal{D}[U]\;  \left[ \det M(\mu_I)\det M(-\mu_I)\right]^{\nicefrac{1}{4}}e^{-\beta S_G}}{\int \!\mathcal{D}[U]\;  \left[ \det M(\mu_I)\det M(-\mu_I)\right]^{\nicefrac{1}{4}}e^{-\beta S_G}} \\
	&= \frac{\langle R_{\mu} O\rangle_{\mu_I}}{\langle R_{\mu} \rangle_{\mu_I}}\,,
	\end{aligned}
\end{equation}
with
\begin{equation}
	\label{eq:Rmu}
	R_{\mu} = \left[\frac{\det M(\mu_u) \det M(\mu_d)}{\det M(\mu_I) \det M(-\mu_I)}\right]^{\nicefrac{1}{4}} \in \mathbb{C}\,.
\end{equation}
The reweighting factor in (\ref{eq:Rmu}) is complex as soon as $\mu_u \neq -\mu_d$,
i.e. when there are baryonic contributions $\mu_B \neq 0$ to the quark chemical
potentials\footnote{We write $\mu_I'$ instead of $\mu_I$ because we do not necessarily keep
$\mu_I$ constant in the reweighting process.}
\begin{equation}
	\label{eq:muu-mud}
	\mu_u = \mu_B + \mu_I' \quad \text{and} \quad \mu_d = \mu_B - \mu_I'.
\end{equation}
The complex nature of $R_\mu$ reflects that direct simulations at baryon
chemical potentials are not possible with standard Monte-Carlo methods.

Combining (\ref{eq:rew-mu}) with the $\lambda$-reweighting in (\ref{eq:rew-lambda}),
it is possible to compute expectation values for two-flavor QCD at arbitrary
chemical potentials in terms of the isospin simulations including a pionic source as
\begin{equation}
	\label{eq:rew-full}
	\langle O \rangle_{\mu_u, \mu_d} = \frac{\langle R_{\mu} R_{\lambda} O
	\rangle_{\mu_I,\lambda}}{\langle R_{\mu}R_{\lambda} \rangle_{\mu_I, \lambda}}
\end{equation}
The observables $O$ are measured in the target ensemble (\ref{eq:Z-mu}).
Using the definitions in (\ref{eq:obs-def}) together with the baryon density,
\begin{equation}
	\label{eq:nB}
	\langle n_B \rangle = \frac{T}{V}\frac{\partial \ln Z_{\mu_u,\mu_d}}{\partial \mu_B}\,,
\end{equation}
the analogue of the isospin density, the important observables are given by
\begin{equation}
	\label{eq:obs-mu}	
	\begin{aligned}
	\langle \bar{\psi} \psi \rangle &= \frac{T}{4V} \left \langle \frac{\partial \ln \det M(\mu_u)}{\partial m_{ud}} + \frac{\partial \ln \det M(\mu_d)}{\partial m_{ud}}\right \rangle \\
	\langle n_I \rangle &= \frac{T}{4V} \left \langle \frac{\partial \ln \det M(\mu_u)}{\partial \mu_u} - \frac{\partial \ln \det M(\mu_d)}{\partial \mu_d} \right \rangle \\
	\langle n_B \rangle &= \frac{T}{4V} \left \langle \frac{\partial \ln \det M(\mu_u)}{\partial \mu_u} + \frac{\partial \ln \det M(\mu_d)}{\partial \mu_d} \right \rangle \,.
	\end{aligned}
\end{equation}
Note that in the target ensemble the pion condensate always vanishes due to
the absence of a pionic source in (\ref{eq:Z-mu}). This is in general the case for
$\lambda=0$ and it is the reason why the associated spontaneous symmetry breaking cannot be
observed in simulations at finite volume.

Our plan to obtain valuable information about the phase diagram is to measure
these observables for many different $(\mu_u, \mu_d)$ pairs in the $\mu_B-\mu_I$ plane
using the combined reweighting procedure from (\ref{eq:rew-full}). In principle this
involves recalculating $R_{\mu}$, eq.~(\ref{eq:Rmu}), many times for each value of
$\mu_u$ and $\mu_d$, which would consume a lot of computer time. To circumvent this,
we use a determinant reduction scheme, first presented in \cite{Toussaint:1989fn},
to express the $(\mu_u,\mu_d)$-dependence of the fermion determinants in a closed
analytic formula,
\begin{equation}
	\label{eq:det-red}
	\begin{aligned}
	\det M(\mu) &= e^{-3V N_t\mu} \det \left(P - e^{N_t\mu}\right)\\
				&= e^{-3V N_t\mu} \prod_{i=1}^{6V}\left(p_i - e^{N_t\mu}\right)\,.
	\end{aligned}
\end{equation}
It is then sufficient to calculate the eigenvalues $p_i$ of the matrix $P$ just once
per configuration. A detailed explanation how to construct $P$ can be found in
ref~\cite{Fodor:2001pe}.

Using this determinant reduction, the computation of the observables in
eq.~(\ref{eq:obs-mu}) reduces to computing the derivatives of $\ln\det M$ with respect to $m_{ud}$
numerically and the ones involving the chemical potentials as
\begin{equation}
	\label{eq:det-der}
	\frac{\partial \ln \det M (\mu)}{\partial \mu} = -3N_tV - N_t \sum_{i=1}^{6V} \frac{e^{N_t\mu}}{p_i - e^{N_t\mu}}\,,
\end{equation}
which follows directly from (\ref{eq:det-red}). For the numerical derivatives
$\partial f(m) / \partial m = \lim_{\Delta m \to 0}[f(m + \Delta m) - f(m)]/\Delta m$
we found the most stable behaviour for $\Delta m \approx 10^{-3} m_{ud}$.

\begin{figure}[t]
	\centering
        \begin{minipage}{.48\textwidth}
        \centering
	\includegraphics[width=\textwidth,clip]{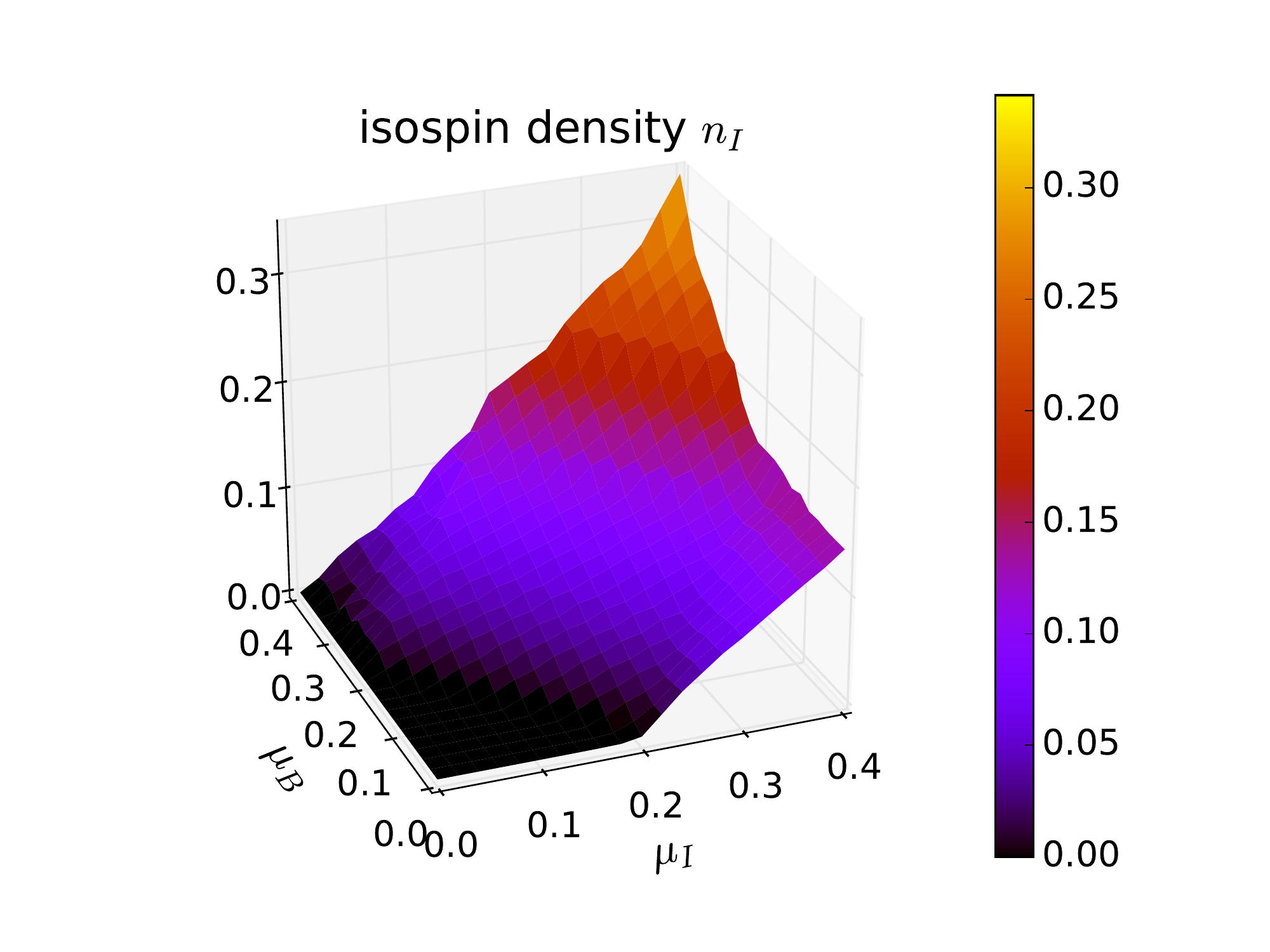}
	\end{minipage}
        \begin{minipage}{.48\textwidth}
        \centering
	\includegraphics[width=\textwidth,clip]{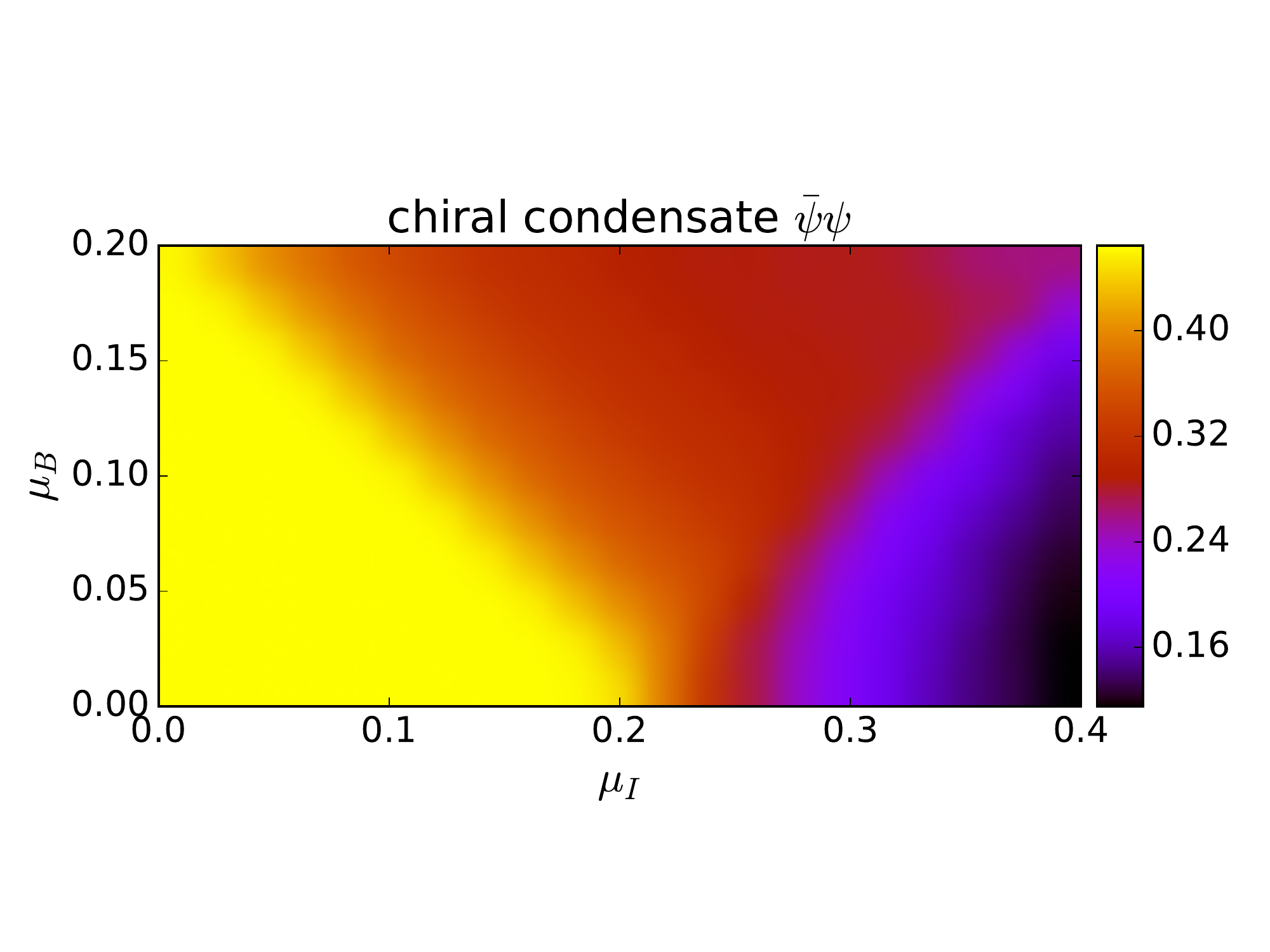}
	\end{minipage}
	\caption{Left: Isospin density $\langle n_I \rangle$ reweighted into the
	$\mu_B-\mu_I$ plane. Right: Projection of the reweighted (non-renormalised)
	chiral condensate $\langle \bar\psi\psi \rangle$ onto the $\mu_B-\mu_I$ plane.}
	\label{fig:rew-nI}
\end{figure}

We show the application of the above machinery to the isospin density in
fig.~\ref{fig:rew-nI}. We simulated at different values of the isospin chemical potential
$\mu_I \in \{ 0.0, 0.1, 0.15, 0.18, 0.2, 0.22, 0.25, 0.3, 0.4 \} $ at $\lambda=0.0025$ and
performed reweighting as in (\ref{eq:rew-full}) to an equidistant grid of $\mu_u$ and
$\mu_d$ values. In order to adress the overlap problem, we chose the auxiliary ensemble such that the error on the reweighted observable is minimal. 

The reweighted isospin density is observed to become nonzero for $\mu_B>m_\pi/2-\mu_I$, i.e.\ outside the triangle around the origin in the left panel of Fig. 10. The reweighted chiral condensate also changes markedly beyond the hypotenuse of this triangle, see the right panel of the same figure (note that here we did not include the renormalization factors from eq.(7)). The fact that this region is connected to the $\mu_B>m_\pi/2$ line at $\mu_I=0$, where the sign problem is known to be severe, calls for a critical interpretation of this tendency.
%
We are currently increasing the statistics to see whether this behaviour remains and employ different methods to estimate the overlap.

Since we cannot access the pion condensate, we determine the phase boundary of the
pion condensation phase indirectly via the chiral condensate.
Since the chiral condensate is expected to decrease at the boundary of the pion
condensation phase, a behaviour also seen in the simulations for $\mu_B=0$, we can
follow the line of the pion condensation phase by searching for the region where this
decrease is visible in the figure. The chiral condensate decreases strongly in the
darker region on the right of the plot and one can clearly see a bend towards larger
values of $\mu_I$ with increasing $\mu_B$.

\section{Conclusions}
\label{sec-4}

In this proceedings article we have presented the current status of our study
of QCD at finite isospin chemical potential with improved staggered fermions
at physical quark masses. The most crucial step in the analysis, the
extrapolation of the prefactor of the pionic source term $\lambda\to0$, has
been done using the improvement scheme introduced in
ref.~\cite{Brandt:2016zdy}, which we have briefly sketched in
sec.~\ref{sec:obs+lambda}.

We have presented new results for the phase diagram on $N_t=8$ and 10 lattices,
supplementing the results for $N_t=6$ presented already in
ref.~\cite{Brandt:2016zdy}. In the approach to the continuum the qualitative
features of the phase diagram remain unchanged. We observe pion condensation
at small temperatures starting at $\mu_I=m_\pi/2$ up to a temperature
of about 90\% of the crossover temperature ($T_c(\mu_I=0)\approx155$~MeV in
the continuum), where the phase boundary starts to shift towards larger values of
$\mu_I$, until it flattens out at around $1.1T_c(0)$. The crossover line shows
a slight downward curvature for smaller values of $\mu_I$. For $\mu_I>m_\pi/2$
we are currently working on the determination of $T_c$ via the inflection
point of the condensate. By comparing the behaviour of the condensate to the
location of the pion condensation phase boundary we obtain first evidence
that the chiral symmetry restoration temperature is consistent with the pion
condensation phase boundary starting from $\mu_I\gtrsim 0.6 m_\pi$.
Concerning the comparison to Taylor expansion around $\mu_I=0$, we have
extended our range in chemical potential for temperatures above the pion condensation
phase boundary and found the range of applicability of the expansion to $O(\mu_I^3)$
to increase with increasing temperature, in agreement with the expectations. Due to
the increase in precision from the improved $\lambda$-extrapolations, we have found
that for $T<T_c$ Taylor expansion to $O(\mu_I)$ fails to describe the data at
$\mu_I$ between 50 to 60~MeV, while the data remains consistent with the expansion
to $O(\mu_I^3)$ up to the phase boundary.

We have also presented first results for the pressure evaluated at finite
isospin chemical potential on $N_t=6$ lattices, which is the first step for our
measurements of the full equation of state. In sec.~\ref{sec:star} we discussed
an application of our measurements of the pressure in terms of the
construction of gravitationally stable pion stars and showed first,
preliminary results for the resulting mass-radius relation. The results
presented so far are unrealistic in the sense that the star is highly charged
and the next mandatory step is the inclusion of further charged particles to
obtain a neutral star.

For an exploration of the phase diagram in the $\mu_B-\mu_I$ plane, we used a
two-step reweighting procedure in the pionic source and the quark chemical
potentials. We have shown the applicability of the leading order approximation
for the reweighting in the pionic source for our test systems, which helps
to reduce the computational cost of the reweighting procedure tremendously.
We provided first results for small baryon chemical potential and obtained
first evidence that the pion condensation phase boundary shifts towards
larger values of the isospin chemical potential for increasing baryon
chemical potential.

\clearpage

\section*{Acknowledgements}

We are grateful to Szabolcs Bors\'anyi for providing the data for the Taylor
expansion coefficients and to Eduardo Fraga, Maur\'icio Hippert and J\"urgen
Schaffner-Bielich for illuminating discussions.
The simulations have been performed on the GPU cluster
of the Institute for Theoretical Physics at the University of Regensburg and
on the FUCHS cluster at the Center for Scientific Computing of the Goethe
University of Frankfurt. The research has been funded by the DFG via the
Emmy Noether Programme EN 1064/2-1 and SFB/TRR 55. B.B. has also received
support from the Frankfurter F\"orderverein f\"ur Physikalische
Grundlagenforschung. S.S. acknowledges support by the Helmholtz Graduate School for Hadron and Ion Research.

\bibliography{lattice2017}

\end{document}